\documentclass[aps,prl,twocolumn,superscriptaddress,floatfix]{revtex4}
\usepackage{graphicx,amsmath}
\usepackage[utf8]{inputenc}

\begin{document}

\title{Many-Body Force and Mobility Measurements in Colloidal Systems}
\author{Jason W. Merrill}
\affiliation{Yale University, Department of Physics}
\author{Sunil K. Sainis}
\affiliation{Yale University, Department of Mechanical Engineering}
\altaffiliation{Now at E Ink, Cambridge Massachusetts }
\author{Jerzy B\l awzdziewicz}
\affiliation{Yale University, Departments of Mechanical Engineering and Physics}
\author{Eric R. Dufresne}
\email[]{eric.dufresne@yale.edu}
\affiliation{Yale University, Departments of Mechanical Engineering, Chemical Engineering, Physics, and Cell Biology}

\date{\today}

\begin{abstract}
We demonstrate a technique for simultaneously measuring each component of the force vectors and mobility tensor of a small collection of colloidal particles based on observing a set of particle trajectories. For a few-body system of micron-sized polymer beads in oil separated by several particle radii, we find that the mobility tensor is well-described by a pairwise Stokeslet model. This stands in contrast to the electrostatic interactions, which were found to deviate significantly from a pairwise model. The measurement technique presented here should be simple to extend to systems of heterogeneous, non-spherical particles arranged in arbitrary 3D geometries.
\end{abstract}

\pacs{}
\maketitle

Typical approaches to understanding the macroscopic behavior of colloidal materials are based on microscopic models including thermal fluctuations and the interactions of pairs of particles. However, collective behaviors of many particles are fundamentally multi-coordinate phenomena that can only be reduced to pair interactions in special circumstances. To that end, techniques for measuring forces in multi-coordinate systems are needed.

Most force measurements are based on Hooke's law; forces of interest are inferred by measuring the deflection of carefully calibrated springs. In the microscopic domain, the surface forces apparatus (SFA) and atomic force microscope (AFM) use mechanical cantilevers \cite{israelachvili_intermolecular_1992}. Static optical tweezers (OT) measurements use an optical potential. These techniques allow sensitive measurements of interactions between pairs of surfaces, but SFA and AFM are all but impossible to extend to multi-particle force measurements. Static optical tweezers can be used to simultaneously measure forces between sets of particles \cite{brunner_direct_2004,mejean_multiplexed_2009}, but such measurements require careful calibration of many traps and may be confounded by optical interactions between particles induced by the tweezers themselves.

Hooke's law is not the only way to measure a force; forces may also be inferred from the trajectories of moving particles. If we wanted to measure the force of gravity on a baseball, we could weigh it on a scale, but we could also toss it in the air and measure its trajectory. Astronomers and particle physicists infer forces from trajectories because appropriate calibrated springs are not available. Such dynamical force measurements have the advantage of not requiring any carefully calibrated springs and avoid perturbing the system being studied. Crocker and Grier introduced a dynamical interaction measurement suitable for colloidal systems called Markovian Dynamics Extrapolation (MDE) \cite{crocker_microscopic_1994}. In this technique, data are collected by first positioning particles in some initial configuration using optical tweezers, and then switching the tweezers off and recording the trajectories of the freely interacting particles using video microscopy.

In Ref. \cite{sainis_statistics_2007}, we introduced an alternative technique for extracting forces from similar trajectory data. Whereas the original MDE analysis requires the entire configuration space to be sampled before the interactions in any particular configuration can be inferred, our analysis uses only local information to calculate forces. This advantage becomes crucial when measuring interactions in multi-particle systems because the size of the configuration space explodes as the number of particles is increased. We have recently used this technique to investigate surfactant-controlled electrostatic interactions between pairs of colloidal particles in a non-polar solvent \cite{sainis_electrostatic_2008}, and deviations from pairwise additivity for the forces between sets of several particles in the same system \cite{merrill_many-body_2009}. These measurements have implicitly involved several coordinates---even a system of two spheres moving in two dimensions has a total of four coordinates---but for ease of presentation, we have previously projected the measured forces onto a single coordinate. Here, we describe the extension of the technique to multiple coordinates and exhibit the full force and mobility curves for each component of a multi-particle system.

The central result of Ref. \cite{sainis_statistics_2007} is a relation between the force, $f$, and the two statistical parameters that describe ensembles of short-time Brownian trajectories, the drift velocity, $v_d$, and the diffusion coefficient, $D$:
\begin{equation}
	f = k_B T D^{-1} v_d.
	\label{eq:force-measurement}
\end{equation}
This is the correct result for a single particle in 1D subject to an external field. As demonstrated in Ref. \cite{sainis_statistics_2007}, this scalar equation also holds separately for each of the hydrodynamic normal modes of a multi-coordinate system, with each normal mode having it's own diffusion coefficient.

To derive this relation, we start with the expression for the drift velocity of a particle in a heavily over-damped viscous medium:
\begin{equation}
	v_d = b f.
	\label{eq:drift-velocity}
\end{equation}
where $b$ is the particle mobility. If the particle mobility were already known, then this relation would be sufficient to determine the forces from particle trajectories simply by measuring $v_d$. In principle, the mobility can be calculated from the properties of the particles and the solvent. Alternatively, the mobility can be eliminated from Eq. (\ref{eq:drift-velocity}) using the Einstein relation:
\begin{equation}
	b = \beta D.
	\label{eq:einstein}
\end{equation}
A simple rearrangement then gives Eq. (\ref{eq:force-measurement}). Since $D$ can be measured directly from the particle trajectories, this eliminates the need to know the mobility ahead of time.

\begin{figure}[t]
\includegraphics[width=.2\columnwidth]{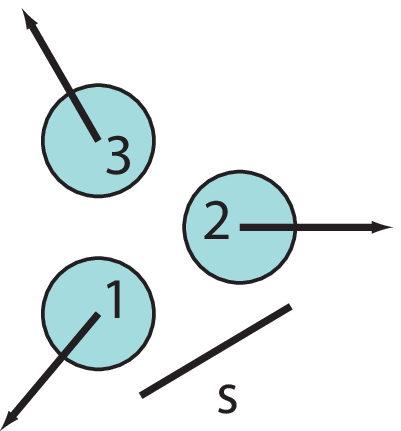}
\includegraphics[width=\columnwidth]{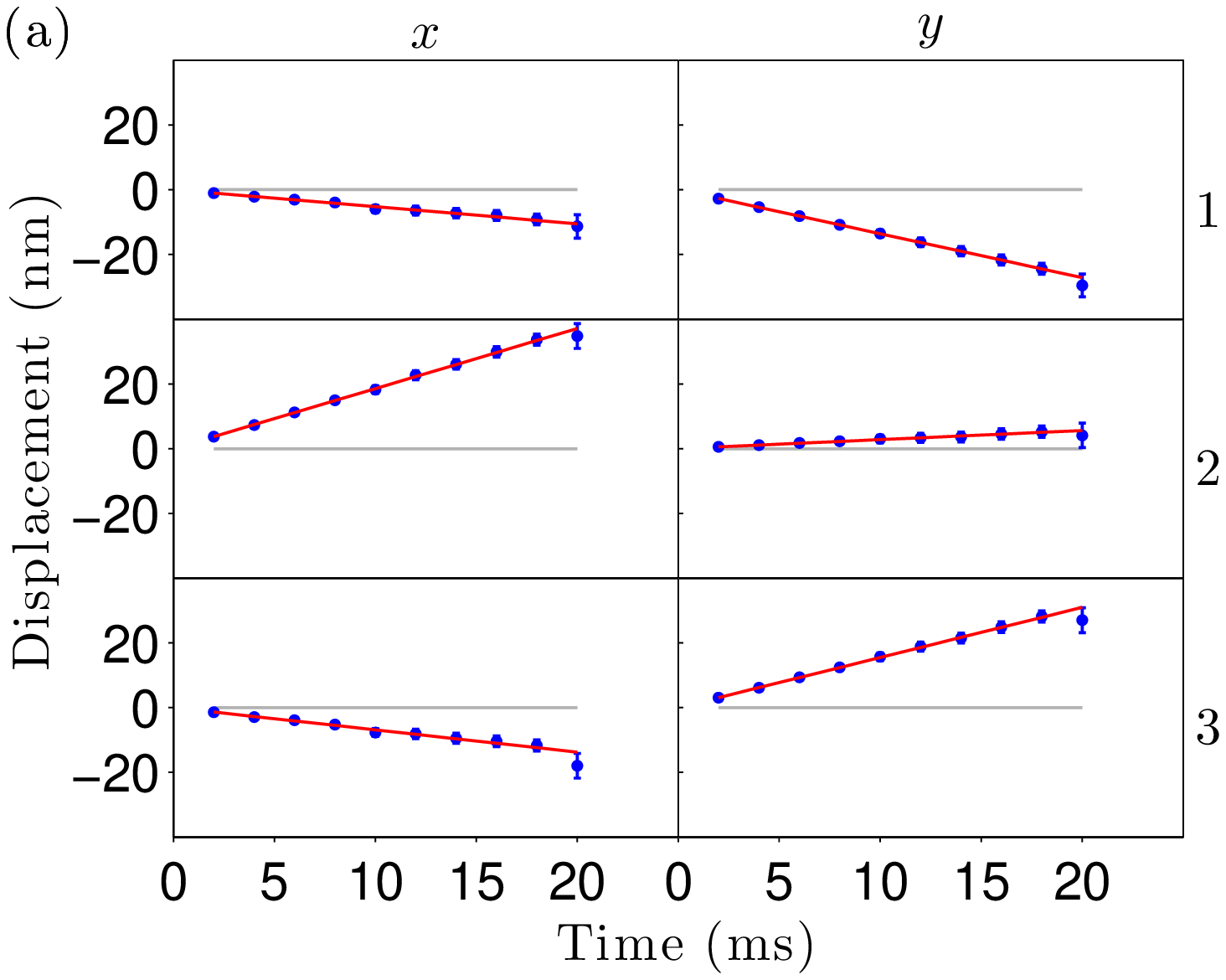}
\includegraphics[width=\columnwidth]{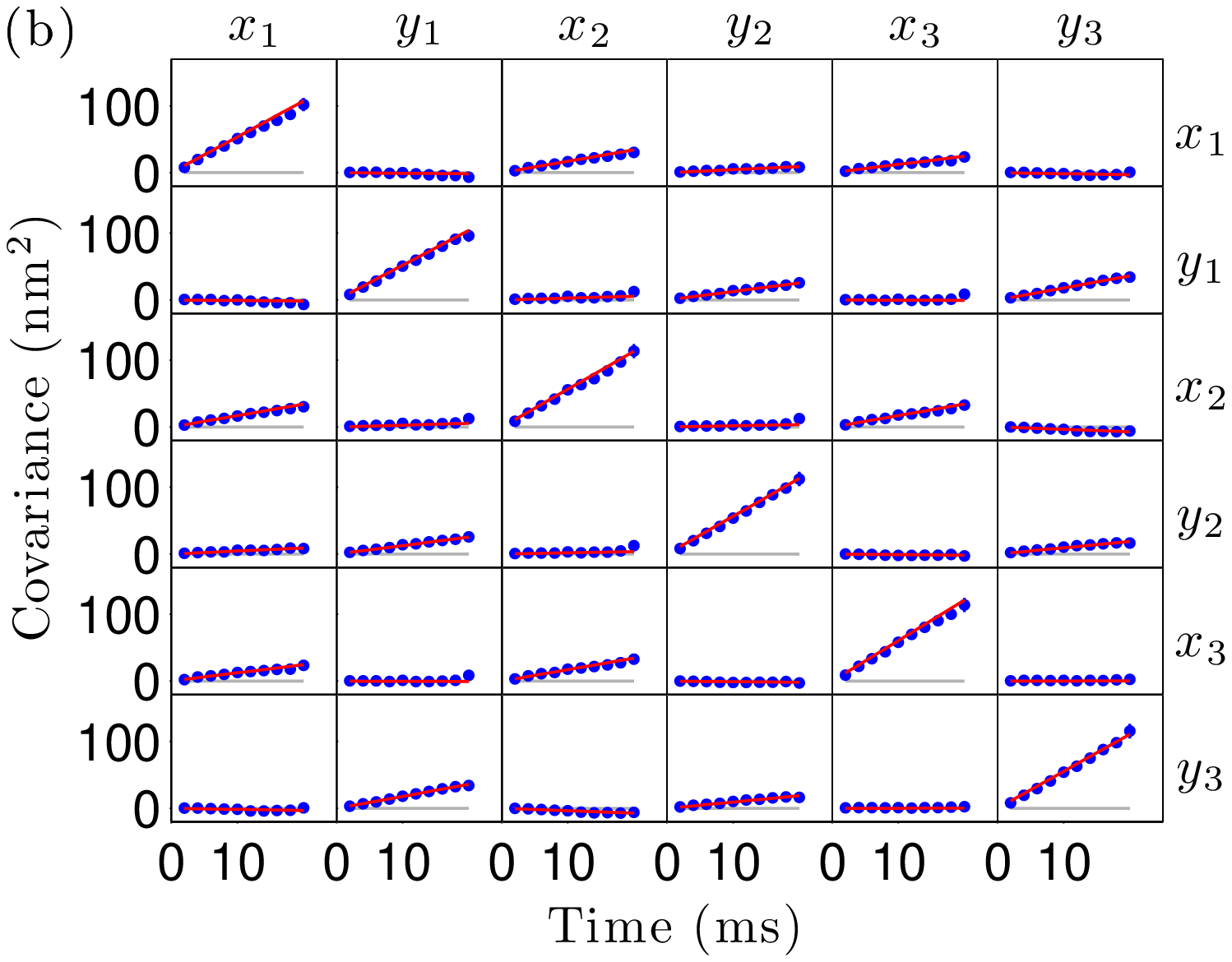}
\caption{\label{fig:line-fitting} \emph{Mean Displacement and Displacement Covariance} (a) Mean displacement and (b) displacement covariance versus time for each coordinate of three particles arranged in an equilateral triangle as shown with side length $s = 4.4 a = 2.6\ \mu$m. Red lines are best fits of Eqns. (\ref{eq:md}) and (\ref{eq:msd}) to the data.}
\end{figure}

This approach is readily generalized to a system of many particles in multiple dimensions. In that case, hydrodynamic interactions lead the drift velocity of one particle to be coupled to the force on all the others. The drift velocity and force vectors, $\mathbf{v}_d$ and $\mathbf{f}$, are linearly related through the mobility tensor, $\mathbf{\tilde{b}}$:
\begin{equation}
	\mathbf{v}_d = \mathbf{\tilde{b}} \mathbf{f}.
	\label{eq:drift-velocity-vector}
\end{equation}
The vector drift velocity and force are made up of the concatenation of the components of the velocity and force for each particle, \textit{e.g.} $\mathbf{v}_d = \{ v_{d,x_1}, v_{d,y_1}, v_{d,x_2}, v_{d,y_2}, ..., v_{d,x_n}, v_{d,y_n} \}$ and similarly for the force vector. Hydrodynamic interactions lead to off-diagonal terms in the mobility tensor, so that a force on one particle may produce a velocity on another particle, or a force directed along one coordinate of a particle may produce a velocity along a different coordinate \cite{kim_microhydrodynamics:_2005}.

A multi-coordinate Einstein relation holds for the mobility and diffusion tensors \cite{kubo_fluctuation-dissipation_1966, batchelor_brownian_1976}:
\begin{equation}
	\mathbf{\tilde{b}} = \beta \mathbf{\tilde{D}},
	\label{eq:einstein-vector}
\end{equation}
so we can write down a multi-coordinate version of Eq. (\ref{eq:force-measurement}) :
\begin{equation}
	\mathbf{f} = k_B T \mathbf{\tilde{D}}^{-1} \mathbf{v}_d.
	\label{eq:force-measurement-vector}
\end{equation}

Since the mobility tensor is symmetric, it can be diagonalized by a set of normal modes. The scalar Eq. (\ref{eq:force-measurement}) holds separately for the force on each normal mode of the mobility tensor, as was shown in Ref. \cite{sainis_statistics_2007}. However, it isn't necessary to solve for the normal modes in order to calculate each component of the force on a multi-particle system. Eq. (\ref{eq:force-measurement-vector}) can be evaluated directly. Contrary to the assumptions of Ref. \cite{sainis_statistics_2007}, there is no need for the force to be directed along a normal mode, nor is there any restriction to conservative forces.

Over appropriate time scales, the mean displacement and displacement covariance increase linearly in time, with slopes related to the components of the drift velocity, $v_{d,i}$, and diffusion tensor, $D_{ij}$ \cite{ermak_brownian_1978}:
\begin{equation}
	\langle x_i(t + \tau) - x_i(\tau) \rangle_\tau = (v_{d,i} + \sum_k\partial_k D_{ki}) t,
	\label{eq:md}
\end{equation}
\begin{equation}
	\mathrm{cov}_\tau(x_i(t + \tau) - x_i(\tau),x_j(t + \tau) - x_j(\tau))  = 2 D_{ij} t + \epsilon_{ij}.
	\label{eq:msd}
\end{equation}
A subtle point here is that the increase of the mean displacement with time given by Eq. (\ref{eq:md}) is due not only to the drift velocity, but also depends on the local gradient of the diffusion tensor.
This correction will become important when particle surfaces are sufficiently close to each other, but it is negligible in the data presented here.

\begin{figure}[t]
\includegraphics[width=.2\columnwidth]{figures/equilateral-configuration.eps}
\includegraphics[width=\columnwidth]{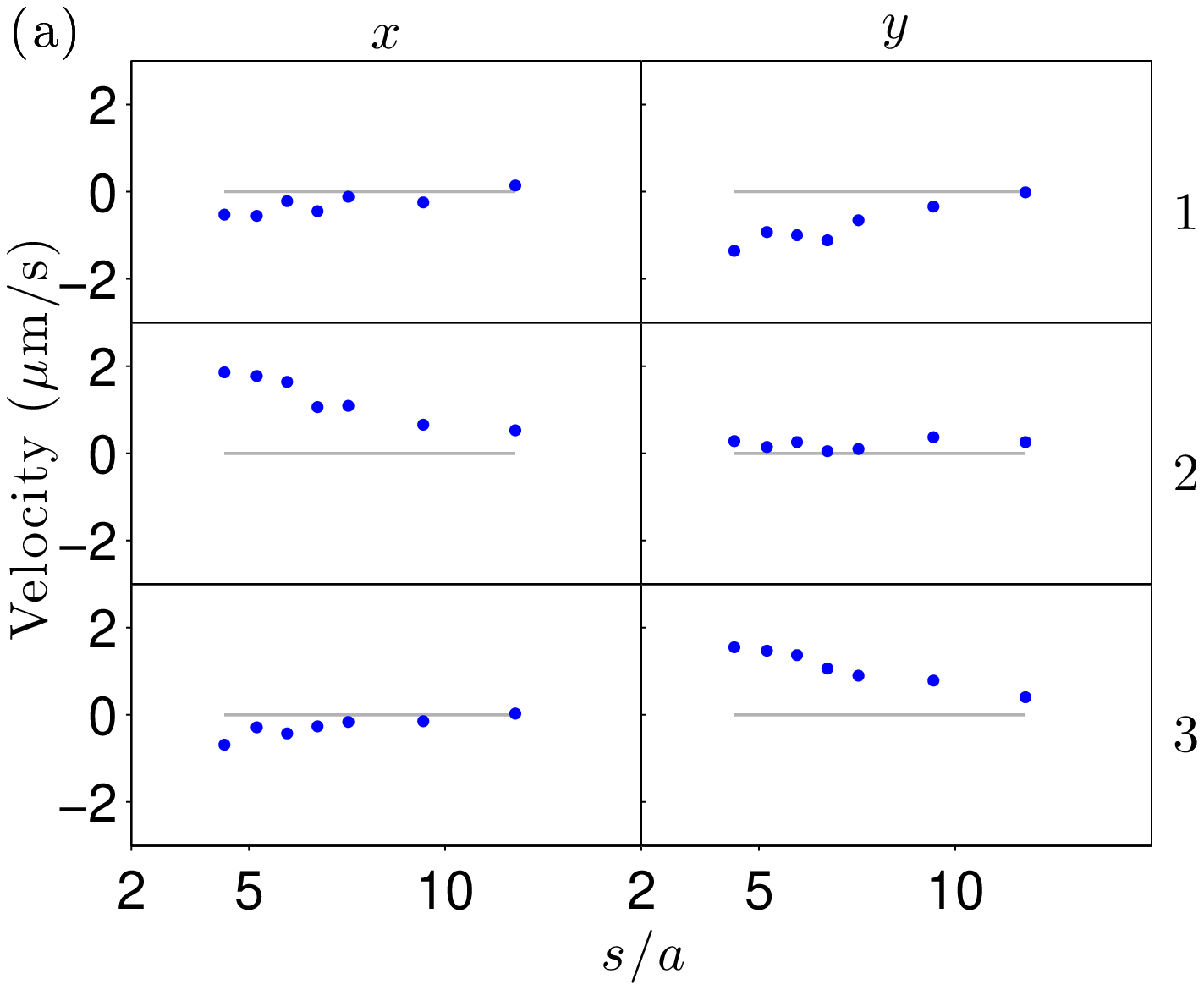}
\includegraphics[width=\columnwidth]{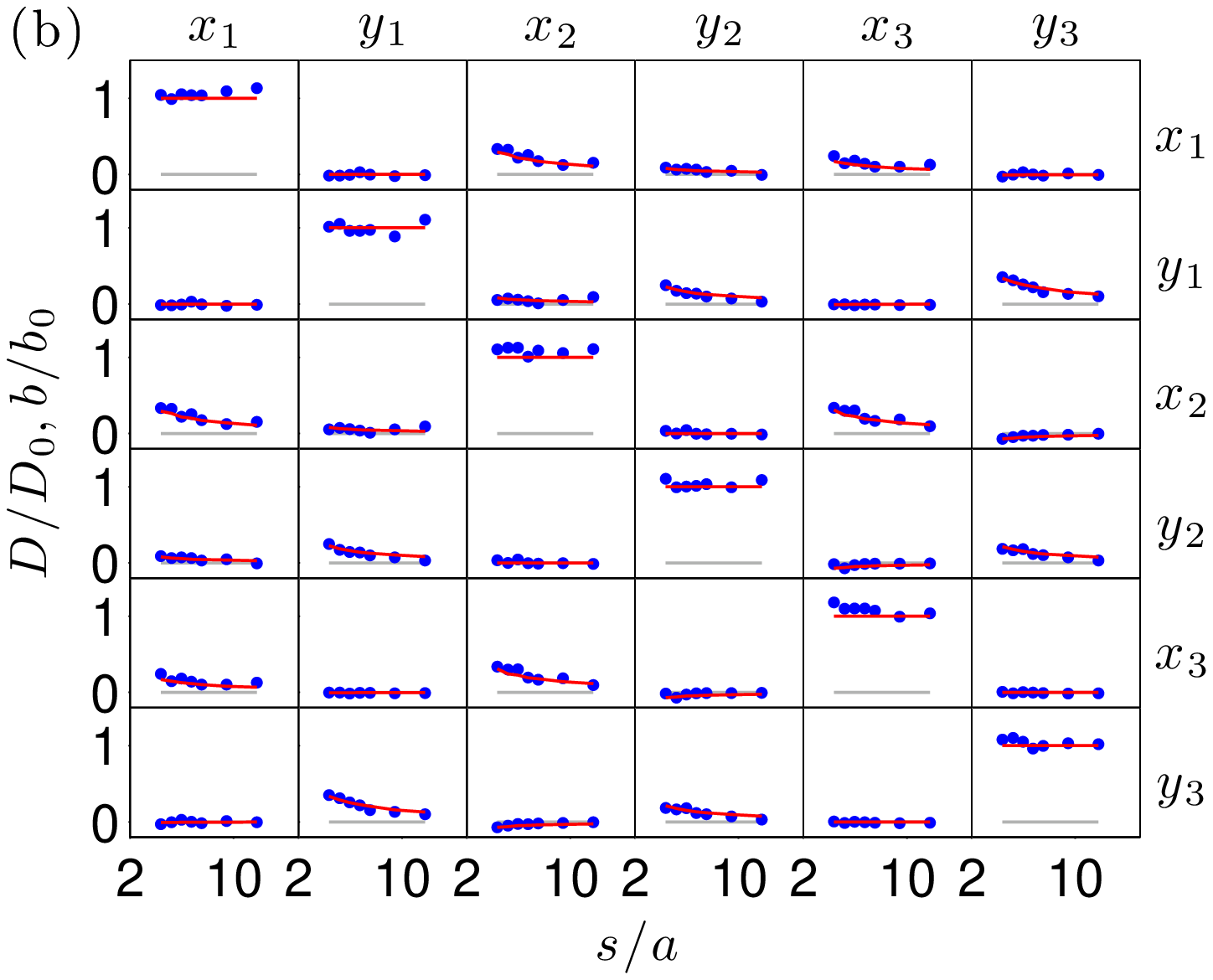}
\caption{\label{fig:v-and-D-vs-s} \emph{Velocity and Diffusion} (a) Drift velocity vector and (b) diffusion/mobility tensor for particles arranged in an equilateral triangle as a function of side length, $s$. Diffusion (mobility) values are normalized to $D_0 = k_B T/6 \pi \eta a = 117$ nm$^2$ms$^{-1}$ ($b_0 = 1/6 \pi \eta a = 29.5$ $\mu$m\,s$^{-1}$pN$^{-1}$). Red lines on the diffusion/mobility tensor plot are predictions based on Eqns. (\ref{eq:self-mobility})  and (\ref{eq:cross-mobility}).}
\end{figure}

This suggests a straightforward approach to measuring the force and mobility in multi-particle systems. We measure the velocity and diffusion by fitting these lines to the statistics of observed trajectories, taking averages for a given delay time, $t$, over the set of non-overlapping pairs of sample times separated by $t$. The intercept term in the displacement covariance equation, $\epsilon_{ij}$, arises due to a combination of uncertainty in position measurements and finite shutter speed \cite{savin_static_2005}.

In the following, we apply this technique to measure the force and mobility for sets of three particles arranged in an equilateral triangle and for seven particles arranged in a hexagon with a particle at its center. The particles are poly(methyl methacrylate) spheres of radius $a = 600$ nm suspended in a solution of 500 $\mu$M NaAOT in hexadecane \cite{sainis_electrostatic_2008}.

The mean displacement and displacement covariance versus time corresponding to Eqns. (\ref{eq:md}) and (\ref{eq:msd}) are shown in Fig. \ref{fig:line-fitting} for particles arranged in an equilateral triangle with side length $s = 4.4 a = 2.6\ \mu$m. This procedure is repeated at a series of side lengths, $s$. The measured drift velocity and diffusion tensor are shown in Fig. \ref{fig:v-and-D-vs-s} for particles arranged in an equilateral triangle with a series of side lengths, $s$.

\begin{figure*}[htb]
\includegraphics[width=.48\textwidth]{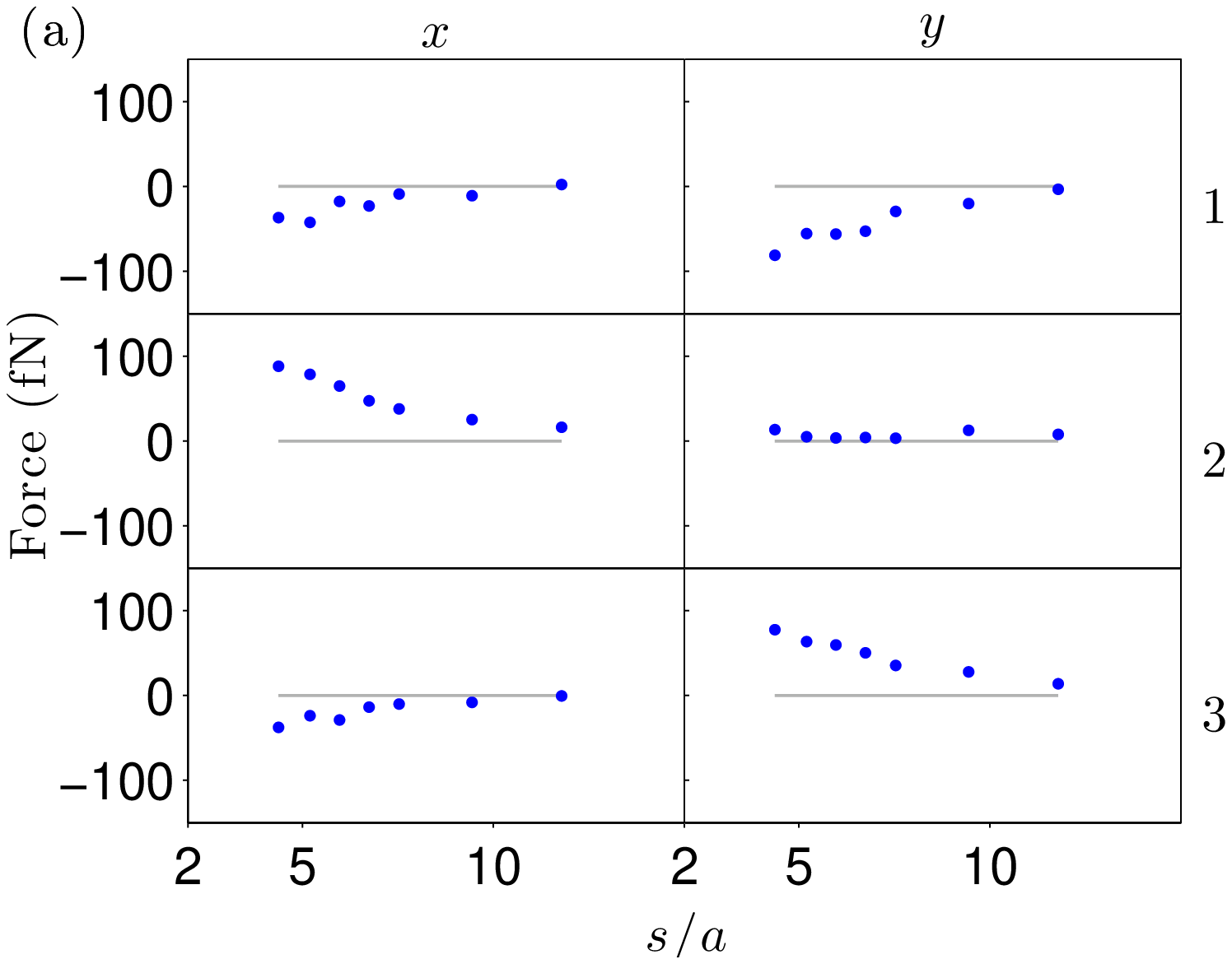}
\includegraphics[width=.48\textwidth]{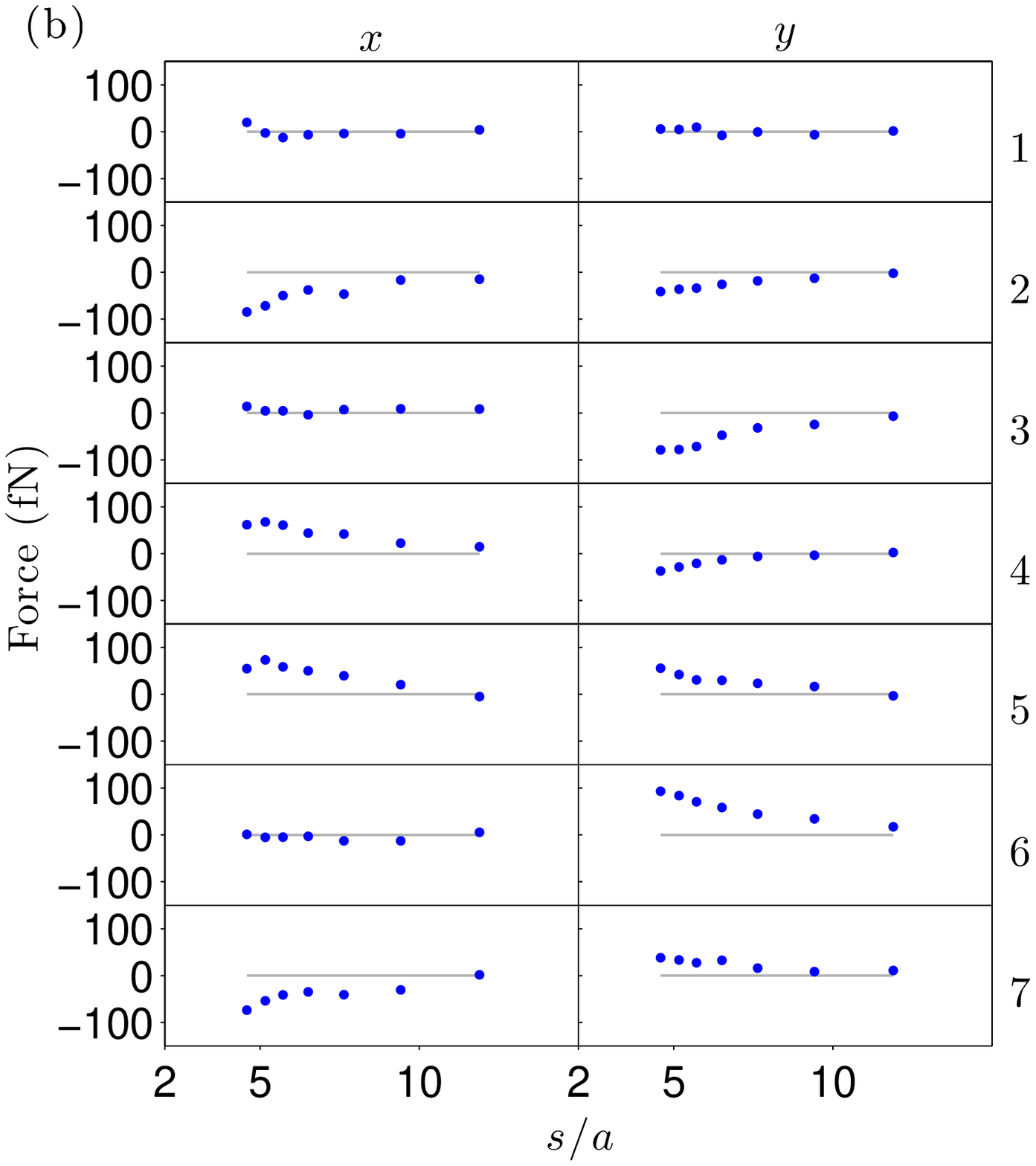}
\includegraphics[width=.1\textwidth]{figures/equilateral-configuration.eps}
\includegraphics[width=.15\textwidth]{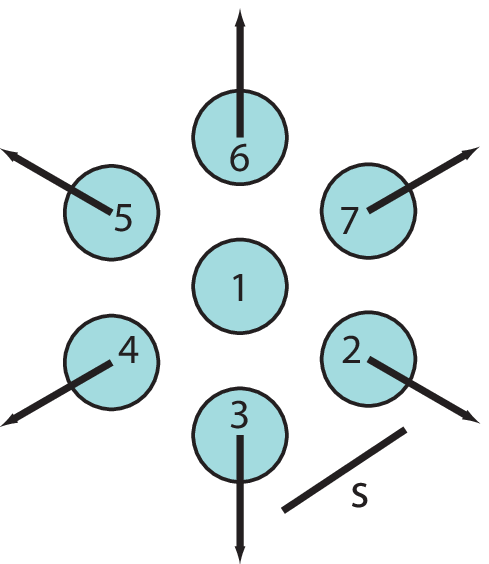}
\caption{\label{fig:f-vs-s} \emph{Forces versus Separation} Components of the force as a function of side length for (a) each of three particles arranged in an equilateral triangle with side length $s$ and (b) seven particles arranged in a hexagon with a particle at its center.}
\end{figure*}

With both $\mathbf{v}_d$ and $\mathbf{\tilde{D}}$ in hand, calculating the entire force vector is a simple application of Eq. (\ref{eq:force-measurement-vector}). The full force vector is shown for a series of separations in Fig. \ref{fig:f-vs-s} for both the equilateral and hexagonal particle configurations. These are the same data as in Ref. \cite{merrill_many-body_2009}, where we reported only forces on the breathing mode calculated using Eq. (\ref{eq:force-measurement}).

There is nothing that limits this analysis to 2D. If 3D position data were available, it could be taken into account with no additional complications. In fact, it would generally be necessary to use 3D data in order to correctly infer the interparticle forces \cite{squires_like-charge_2000}; however, if all the particles are in plane and the surroundings are isotropic, the force measurement splits as
\begin{eqnarray}
	\mathbf{f}_{xy} & = &  k_B T \mathbf{\tilde{D}}_{xy}^{-1} \mathbf{v}_{d,xy}, \\
	\mathbf{f}_z & = &  k_B T \mathbf{\tilde{D}}_z^{-1} \mathbf{v}_{d,z}.
\end{eqnarray}
We expect the forces to be central and thus in plane, so we need only observe the in plane position in order to find the forces. The observed trajectories must be short enough that the particles remain roughly in plane. We ensure that the surroundings are effectively isotropic by conducting our experiments more than 30 particle diameters from any wall.

To this point, we have refrained from giving any interpretation for the form of the velocity, diffusion, and force versus separation curves in order to emphasize that our force calculation is independent of any model of the particle interactions. Our only assumptions are that 1) the Einstein relation, Eq. (\ref{eq:einstein-vector}), holds; 2) the drift velocity and diffusion coefficients may be considered locally constant over a single trajectory; and 3) the gradient of the diffusion/mobility tensor can be neglected in determining the drift velocity. We satisfy conditions 1 and 2 by choosing appropriate experimental timescales, \textit{i.e.} the frame rate and number of frames in single trajectory from Eq. (\ref{eq:md}). We can check condition 3 by examining the diffusion tensor data in Fig. \ref{fig:v-and-D-vs-s}. The magnitude of $\mathbf{\nabla} \cdot \tilde{\mathbf{D}}$ is less than 0.1 $\mu$m/s at the smallest measured separation, which is a negligible fraction of the drift velocity. Notably, we have not made any assumption that the forces or hydrodynamic coupling are pairwise. With the present technique, we can \emph{measure} whether the interactions are pairwise.

\begin{figure*}[t]
\includegraphics[width=.47\textwidth]{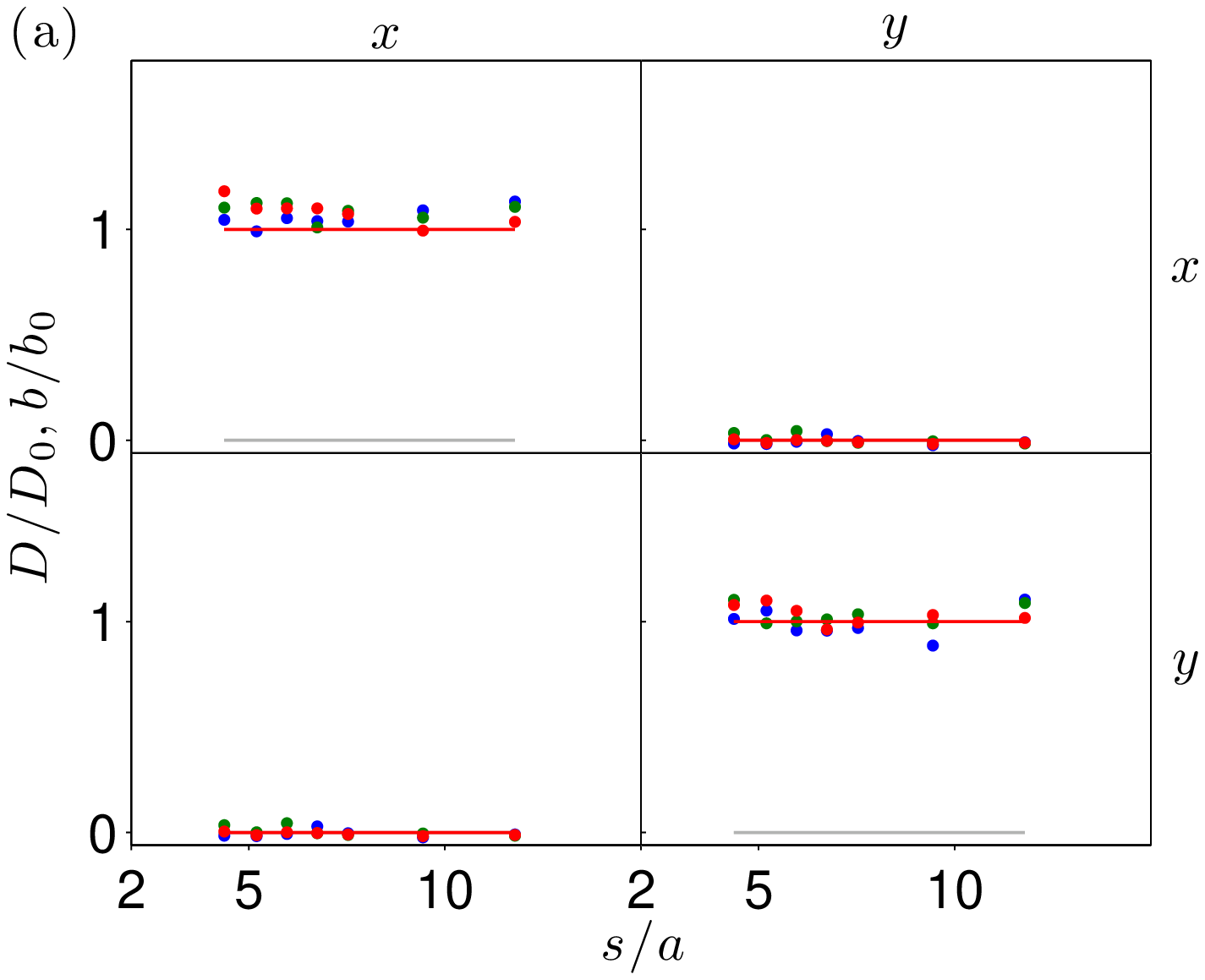}
\includegraphics[width=.49\textwidth]{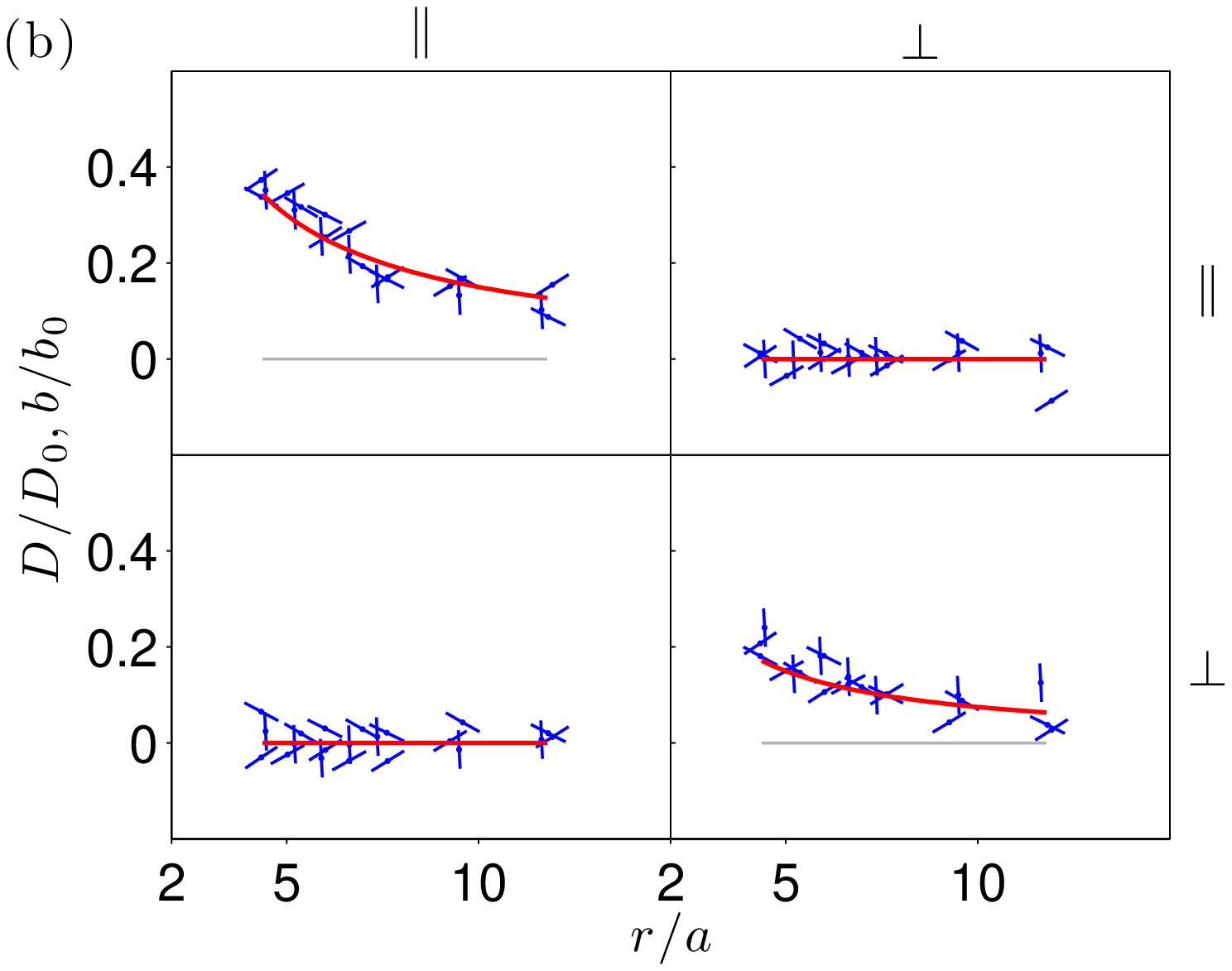}
\includegraphics[width=.47\textwidth]{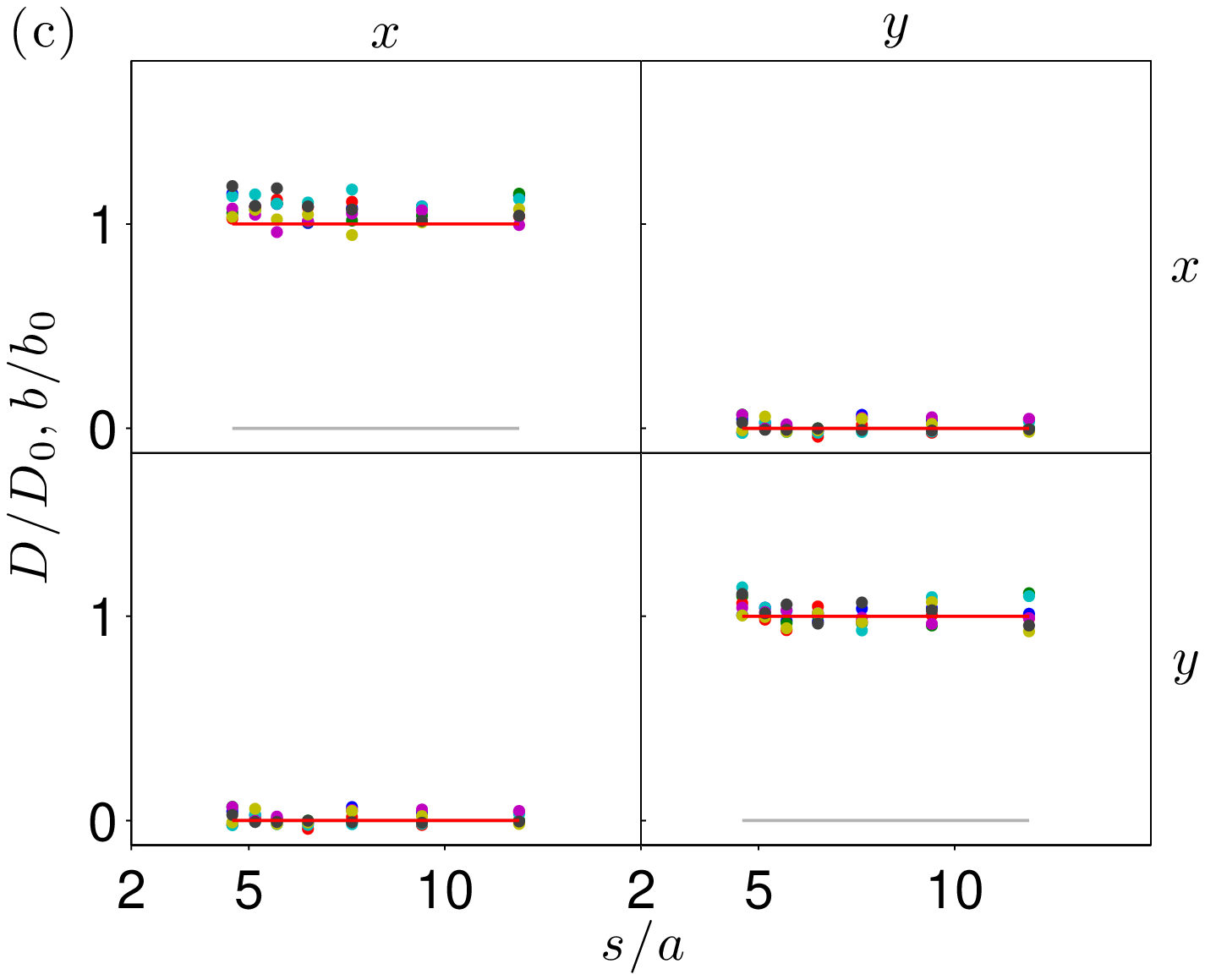}
\includegraphics[width=.49\textwidth]{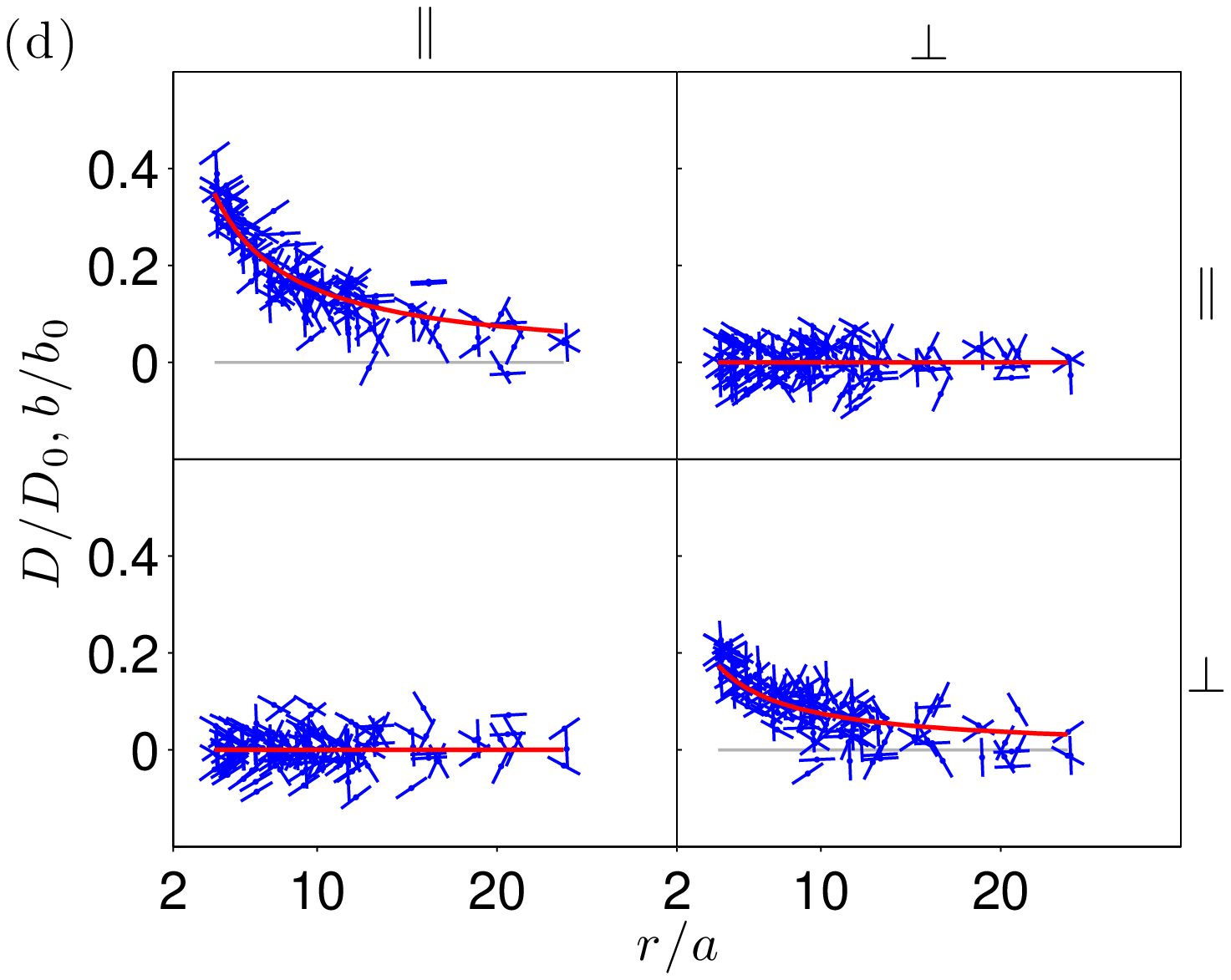}
\caption{\label{fig:reduced-diffusion} \emph{Self-Mobility and Cross-Mobility} Self blocks (a,c) and cross blocks (b,d) of the diffusion/mobility tensor for equilateral (a,b) and hexagonal (c,d) particle configurations. Red lines are the predictions of Eqns. (\ref{eq:self-mobility}) and (\ref{eq:cross-mobility-pair-components}). The self-mobility blocks are expressed in fixed camera coordinates, and the cross-mobility blocks are expressed in pair coordinates as described in the text. The self-mobility blocks reference only a single particle, so the horizontal axis is labeled by the side length, $s$, of the configuration. The cross-mobility blocks reference two particles, so the horizontal axis is labeled by the separation, $r$, between the pair. In the equilateral configuration, all particles are separated by the same distance, so there is no distinction between $r$ and $s$, but in the hexagonal configuration, there are several different pair separations, $r$, for a single side length, $s$. In the cross-mobility plots (b,d), each point has a line through it indicating the direction of the pair separation axis in fixed camera coordinates.}
\end{figure*}

Having established the general force measurement technique, we now explore the structure of the multi-particle diffusion/mobility tensor. The interparticle forces in this data set were interpreted in Ref. \cite{merrill_many-body_2009}, where we found good agreement with linearized Poisson-Boltzmann theory so long as particle surfaces were taken to be at constant electrostatic potential rather than constant charge density. These boundary conditions lead to the observed non-pairwise additivity of the forces.

\begin{figure*}[t]
\includegraphics[width=.49\textwidth]{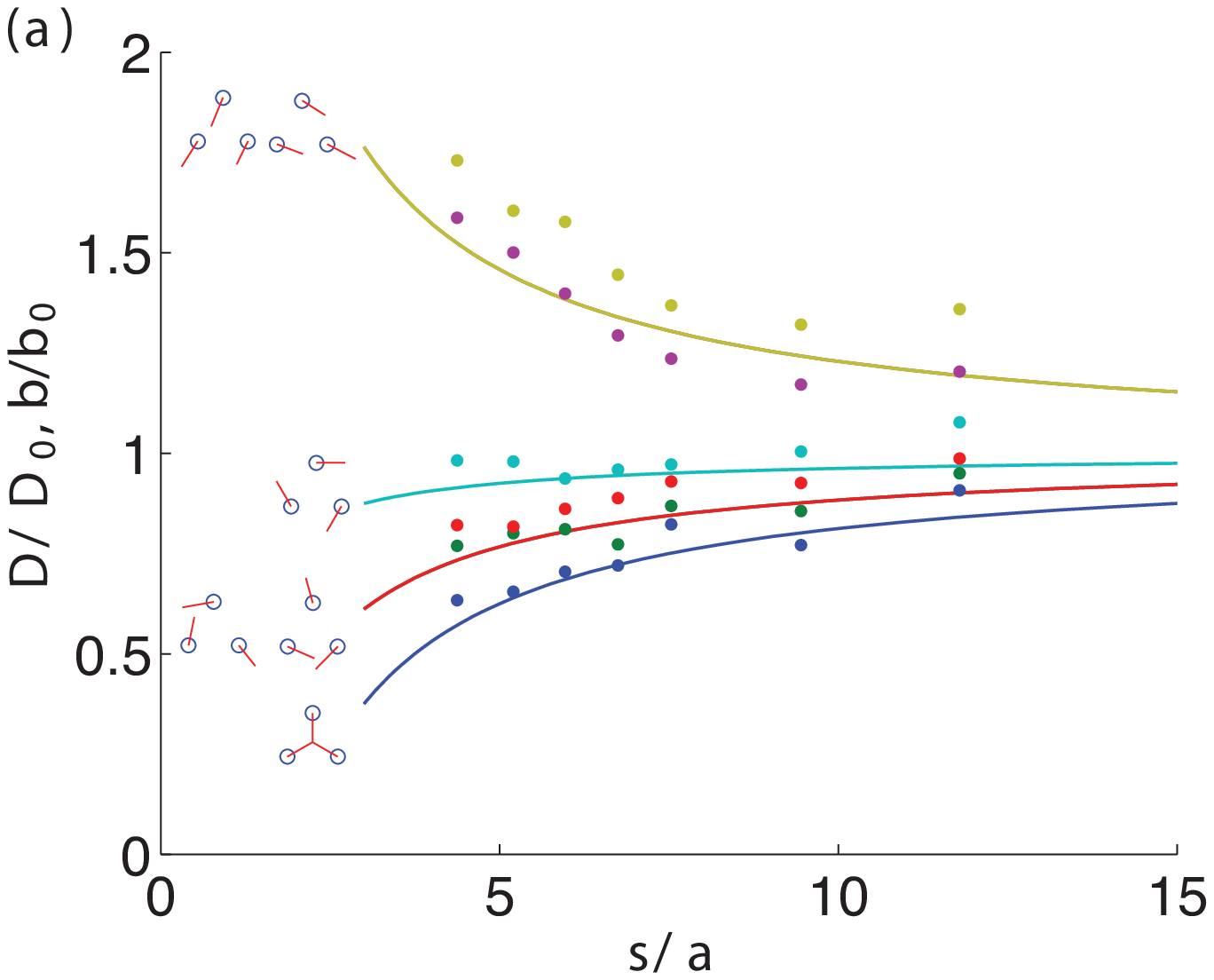}
\includegraphics[width=.49\textwidth]{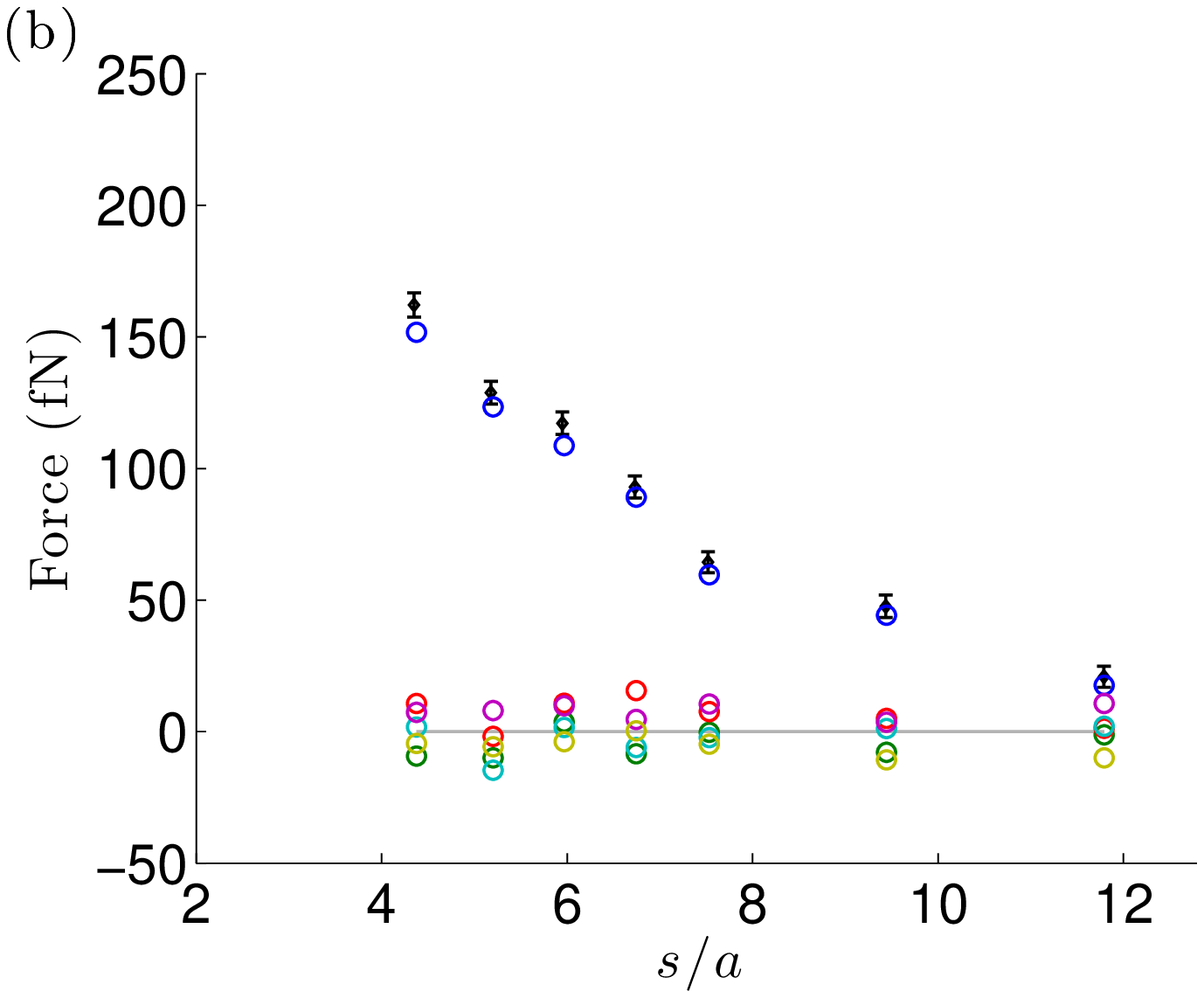}
\caption{\label{fig:normal-mode-diffusion} \emph{Normal mode mobility and projected force} (a) Mobility versus side length for each of the 6 linearly independent normal modes of the equilateral configuration. (b) Force projected onto each of the normal modes. Only the breathing mode has a large non-zero force. Black points with error bars show the result of analyzing the perimeter coordinate with the scalar Eq. (\ref{eq:force-measurement}), as in Ref. \cite{merrill_many-body_2009}.}
\end{figure*}

In these data, where the separation between particles is always greater than 4 particle radii, the mobility data are well-described by the Stokeslet Superposition Approximation (SSA) \cite{pozrikidis_boundary_1992}. In this approximation, the mobility is expanded in powers of $1/r$ and truncated at first order. More accurate mobility calculations for smaller separations are described in Ref. \cite{cichocki_friction_1994}.

The mobility tensor can be split up into particle-particle blocks ($2 \times 2$ in this 2D case). The diagonal blocks represent the self-mobility of a single particle and are given in the SSA by
\begin{equation}
	\tilde{\mathbf{b}}^s = \frac{1}{6 \pi \eta a} \tilde{\mathbf{I}}_2 = b_0 \tilde{\mathbf{I}}_2,
	\label{eq:self-mobility}
\end{equation}
where $\eta$ is the dynamic viscosity of the fluid, $a$ is the radius of the particle, and $\tilde{\mathbf{I}}_2$ is the $2 \times 2$ identity. The off-diagonal blocks represent hydrodynamic interactions between a pair of particles, \textit{i.e.} the cross-mobility, and are given in the SSA by
\begin{equation}
	\tilde{\mathbf{b}}^c = \frac{1}{8 \pi \eta r} (\tilde{\mathbf{I}}_2 +  \hat{\mathbf{r}}\hat{\mathbf{r}}),
    \label{eq:cross-mobility}
\end{equation}
where $r$ is the separation between the particles, and $\hat{\mathbf{r}}\hat{\mathbf{r}}$ is a projection operator that projects the force along the pair separation axis. In the fixed camera basis, its components are
\begin{equation}
	\left[\hat{\mathbf{r}}\hat{\mathbf{r}}\right]_\mathrm{cam.} =
	    \begin{pmatrix} \cos^2(\theta) & \sin(\theta)\cos(\theta) \\
	                    \sin(\theta)\cos(\theta) & \sin^2(\theta) \\
	    \end{pmatrix},
	\label{eq:projection}
\end{equation}
where $\theta$ is the angle between the pair separation axis and the camera $x$-axis. Predictions of this theory based on the nominal particle radius, $a = 600$ nm, are shown as red lines in Fig. \ref{fig:v-and-D-vs-s}, and are in good agreement with the measurement.

In order to further bring out the structure of the diffusion/mobility tensor, we note that for identical particles, in this approximation the self-mobility blocks should all be identical and constant, and the cross-mobility blocks should all collapse when expressed in coordinates adapted to each pair with basis vectors parallel and perpendicular to the separation axis ($\theta = 0$). In these pair coordinates, the components of the cross-mobility blocks are simply
\begin{equation}
	\left[\tilde{\mathbf{b}}^c\right]_\mathrm{pair} = \frac{1}{8 \pi \eta r}
	\begin{pmatrix} 2 & 0 \\ 0 & 1 \end{pmatrix}.
	\label{eq:cross-mobility-pair-components}
\end{equation}

We plot measured values of the self-mobility blocks as a function of side length, $s$, and the cross-mobility blocks as a function of separation, $r$, for both the equilateral and hexagonal configurations in Fig. \ref{fig:reduced-diffusion}. For the cross-mobility blocks, the direction of the interparticle separation is represented by the direction of a blue line through each point. The Stokeslet predictions are plotted as red lines. Agreement is generally quite good, although the diffusion coefficient along the camera's $x$-axis is slightly larger than the diffusion coefficient along the $y$-axis. We believe that this is due to anisotropic fluctuations in the microscope stage, which lead to a small increase in the apparent diffusion coefficient along the $x$-axis. This effect is also visible in the cross-mobility blocks, where a small systematic dependence on the direction of the pair separation axis can be discerned. The consistent collapse of these curves demonstrates that hydrodynamic interactions are essentially pairwise for these separations, in contrast to the electrostatic interactions in this system which showed significant departures from the pairwise model \cite{merrill_many-body_2009}.

Analysis of the normal modes of the diffusion/mobility tensor gives additional insight. From the mobility point of view, when a force is applied along one of these normal modes, it induces a velocity in that normal mode only and does not couple to the other normal modes. From the diffusion point of view, fluctuations in these coordinates are uncorrelated. Fig. \ref{fig:normal-mode-diffusion} shows the mobility as a function of side length of the equilateral triangle for each of the 6 linearly independent normal modes \cite{di_leonardo_eigenmodes_2007}. Two of the normal modes are unique; the others pair to form two degenerate subspaces. The degeneracy is a consequence of the 3-fold rotation symmetry of the equilateral configuration. Rotating any normal mode by $2 \pi/3$ leads to another normal mode with the same eigenvalue. The original mode and the rotated mode will then span a degenerate subspace unless they are in fact the same mode---that is, unless the original mode possesses the full 3-fold symmetry of the particle configuration. Fig. \ref{fig:normal-mode-diffusion}b shows the magnitude of the projection of the total force onto each of the normal modes. Since the particles are nearly identical and the force is central, the force is directed almost entirely along the breathing mode.

Given this observation, it is interesting to consider whether the force could be characterized by reducing the system to 1D at the outset by considering only the perimeter of the equilateral triangle at each time and calculating the force using the scalar Eq. (\ref{eq:force-measurement}). An advantage of this analysis is that the perimeter is insensitive to drifts and fluctuations in the microscope stage. The result of this procedure is shown in Fig. \ref{fig:normal-mode-diffusion} as black points with error bars, as presented in Ref. \cite{merrill_many-body_2009}. In these data, there is good agreement with the forces calculated by projecting at the end, but such agreement should not be taken for granted. When working with non-Cartesian coordinates (like the perimeter), it is generally necessary to make geometrical corrections to Eq. (\ref{eq:md}) that reflect variations in the volume of phase space associated with a given coordinate. The necessary corrections can be derived by expanding the mean and mean square displacements of the perimeter (or other coordinate function) in small displacements around the initial configuration, keeping terms only up to first order in $dt$.

As an example, consider the case of two particles. If there are no forces, then according to Eq. (\ref{eq:drift-velocity-vector}) there should be no drift velocity for any of the Cartesian coordinates of the particles. Nevertheless, we expect diffusion will cause the mean separation between the particles to increase in time. If we consider the separation as a single coordinate and naïvely equate the slope of the mean displacement versus time with the drift velocity, then upon applying Eq. (\ref{eq:force-measurement}) we will find a non-zero ``effective force.'' For the separation, $r$, between two particles, the effective force is $f_\mathrm{eff} = k_B T/r$. For particles separated by 2 microns at room temperature, the effective force is 2 fN, which is just below our experimental resolution.

Measurements using this technique have already provided valuable insight into the charging mechanism of colloids in non-polar solvents. Soon, we hope to examine systems of particles with dissimilar charges and aqueous systems. Such force measurements are not limited to translational coordinates. It should be possible to use the same technique to measure torques on anisotropic particles.

We acknowledge support from the National Institute for Nano Engineering at Sandia National Laboratory, an NSF CAREER grant to E.R.D. (CBET-0547294) and an NSF grant to J.B. and E.R.D. (CBET-0931504).


\begin{thebibliography}{15}
\expandafter\ifx\csname natexlab\endcsname\relax\def\natexlab#1{#1}\fi
\expandafter\ifx\csname bibnamefont\endcsname\relax
  \def\bibnamefont#1{#1}\fi
\expandafter\ifx\csname bibfnamefont\endcsname\relax
  \def\bibfnamefont#1{#1}\fi
\expandafter\ifx\csname citenamefont\endcsname\relax
  \def\citenamefont#1{#1}\fi
\expandafter\ifx\csname url\endcsname\relax
  \def\url#1{\texttt{#1}}\fi
\expandafter\ifx\csname urlprefix\endcsname\relax\def\urlprefix{URL }\fi
\providecommand{\bibinfo}[2]{#2}
\providecommand{\eprint}[2][]{\url{#2}}

\bibitem[{\citenamefont{Israelachvili}(1992)}]{israelachvili_intermolecular_19%
92}
\bibinfo{author}{\bibfnamefont{J.~N.} \bibnamefont{Israelachvili}},
  \emph{\bibinfo{title}{Intermolecular and Surface Forces}}
  (\bibinfo{publisher}{Academic Press}, \bibinfo{address}{New York},
  \bibinfo{year}{1992}).

\bibitem[{\citenamefont{Brunner et~al.}(2004)\citenamefont{Brunner, Dobnikar,
  von Grünberg, and Bechinger}}]{brunner_direct_2004}
\bibinfo{author}{\bibfnamefont{M.}~\bibnamefont{Brunner}},
  \bibinfo{author}{\bibfnamefont{J.}~\bibnamefont{Dobnikar}},
  \bibinfo{author}{\bibfnamefont{H.}~\bibnamefont{von Grünberg}},
  \bibnamefont{and}
  \bibinfo{author}{\bibfnamefont{C.}~\bibnamefont{Bechinger}},
  \bibinfo{journal}{Physical Review Letters} \textbf{\bibinfo{volume}{92}},
  \bibinfo{pages}{078301} (\bibinfo{year}{2004}).

\bibitem[{\citenamefont{Mejean et~al.}(2009)\citenamefont{Mejean, Schaefer,
  Millman, Forscher, and Dufresne}}]{mejean_multiplexed_2009}
\bibinfo{author}{\bibfnamefont{C.~O.} \bibnamefont{Mejean}},
  \bibinfo{author}{\bibfnamefont{A.~W.} \bibnamefont{Schaefer}},
  \bibinfo{author}{\bibfnamefont{E.~A.} \bibnamefont{Millman}},
  \bibinfo{author}{\bibfnamefont{P.}~\bibnamefont{Forscher}}, \bibnamefont{and}
  \bibinfo{author}{\bibfnamefont{E.~R.} \bibnamefont{Dufresne}},
  \bibinfo{journal}{Optics Express} \textbf{\bibinfo{volume}{17}},
  \bibinfo{pages}{6209} (\bibinfo{year}{2009}).

\bibitem[{\citenamefont{Crocker and Grier}(1994)}]{crocker_microscopic_1994}
\bibinfo{author}{\bibfnamefont{J.~C.} \bibnamefont{Crocker}} \bibnamefont{and}
  \bibinfo{author}{\bibfnamefont{D.~G.} \bibnamefont{Grier}},
  \bibinfo{journal}{Physical Review Letters} \textbf{\bibinfo{volume}{73}},
  \bibinfo{pages}{352} (\bibinfo{year}{1994}).

\bibitem[{\citenamefont{Sainis et~al.}(2007)\citenamefont{Sainis, Germain, and
  Dufresne}}]{sainis_statistics_2007}
\bibinfo{author}{\bibfnamefont{S.~K.} \bibnamefont{Sainis}},
  \bibinfo{author}{\bibfnamefont{V.}~\bibnamefont{Germain}}, \bibnamefont{and}
  \bibinfo{author}{\bibfnamefont{E.~R.} \bibnamefont{Dufresne}},
  \bibinfo{journal}{Physical Review Letters} \textbf{\bibinfo{volume}{99}},
  \bibinfo{pages}{018303} (\bibinfo{year}{2007}).

\bibitem[{\citenamefont{Sainis et~al.}(2008)\citenamefont{Sainis, Merrill, and
  Dufresne}}]{sainis_electrostatic_2008}
\bibinfo{author}{\bibfnamefont{S.~K.} \bibnamefont{Sainis}},
  \bibinfo{author}{\bibfnamefont{J.~W.} \bibnamefont{Merrill}},
  \bibnamefont{and} \bibinfo{author}{\bibfnamefont{E.~R.}
  \bibnamefont{Dufresne}}, \bibinfo{journal}{Langmuir}
  \textbf{\bibinfo{volume}{24}}, \bibinfo{pages}{13334} (\bibinfo{year}{2008}).

\bibitem[{\citenamefont{Merrill et~al.}(2009)\citenamefont{Merrill, Sainis, and
  Dufresne}}]{merrill_many-body_2009}
\bibinfo{author}{\bibfnamefont{J.~W.} \bibnamefont{Merrill}},
  \bibinfo{author}{\bibfnamefont{S.~K.} \bibnamefont{Sainis}},
  \bibnamefont{and} \bibinfo{author}{\bibfnamefont{E.~R.}
  \bibnamefont{Dufresne}}, \bibinfo{journal}{Physical Review Letters}
  \textbf{\bibinfo{volume}{103}}, \bibinfo{pages}{138301}
  (\bibinfo{year}{2009}).

\bibitem[{\citenamefont{Kim and Karrila}(2005)}]{kim_microhydrodynamics:_2005}
\bibinfo{author}{\bibfnamefont{S.}~\bibnamefont{Kim}} \bibnamefont{and}
  \bibinfo{author}{\bibfnamefont{S.~J.} \bibnamefont{Karrila}},
  \emph{\bibinfo{title}{Microhydrodynamics: Principles and Selected
  Applications}} (\bibinfo{publisher}{Dover Publications},
  \bibinfo{address}{Mineola, New York}, \bibinfo{year}{2005}).

\bibitem[{\citenamefont{Kubo}(1966)}]{kubo_fluctuation-dissipation_1966}
\bibinfo{author}{\bibfnamefont{R.}~\bibnamefont{Kubo}},
  \bibinfo{journal}{Reports on Progress in Physics}
  \textbf{\bibinfo{volume}{29}}, \bibinfo{pages}{255} (\bibinfo{year}{1966}).

\bibitem[{\citenamefont{Batchelor}(1976)}]{batchelor_brownian_1976}
\bibinfo{author}{\bibfnamefont{G.~K.} \bibnamefont{Batchelor}},
  \bibinfo{journal}{Journal of Fluid Mechanics} \textbf{\bibinfo{volume}{74}},
  \bibinfo{pages}{1} (\bibinfo{year}{1976}).

\bibitem[{\citenamefont{Ermak and {McCammon}}(1978)}]{ermak_brownian_1978}
\bibinfo{author}{\bibfnamefont{D.~L.} \bibnamefont{Ermak}} \bibnamefont{and}
  \bibinfo{author}{\bibfnamefont{J.~A.} \bibnamefont{{McCammon}}},
  \bibinfo{journal}{The Journal of Chemical Physics}
  \textbf{\bibinfo{volume}{69}}, \bibinfo{pages}{1352} (\bibinfo{year}{1978}).

\bibitem[{\citenamefont{Savin and Doyle}(2005)}]{savin_static_2005}
\bibinfo{author}{\bibfnamefont{T.}~\bibnamefont{Savin}} \bibnamefont{and}
  \bibinfo{author}{\bibfnamefont{P.~S.} \bibnamefont{Doyle}},
  \bibinfo{journal}{Biophysical Journal} \textbf{\bibinfo{volume}{88}},
  \bibinfo{pages}{623} (\bibinfo{year}{2005}).

\bibitem[{\citenamefont{Squires and Brenner}(2000)}]{squires_like-charge_2000}
\bibinfo{author}{\bibfnamefont{T.~M.} \bibnamefont{Squires}} \bibnamefont{and}
  \bibinfo{author}{\bibfnamefont{M.~P.} \bibnamefont{Brenner}},
  \bibinfo{journal}{Physical Review Letters} \textbf{\bibinfo{volume}{85}},
  \bibinfo{pages}{4976} (\bibinfo{year}{2000}).

\bibitem[{\citenamefont{Pozrikidis}(1992)}]{pozrikidis_boundary_1992}
\bibinfo{author}{\bibfnamefont{C.}~\bibnamefont{Pozrikidis}},
  \emph{\bibinfo{title}{Boundary Integral and Singularity Methods for
  Linearized Viscous Flow}} (\bibinfo{publisher}{Cambridge University Press},
  \bibinfo{year}{1992}).

\bibitem[{\citenamefont{Cichocki et~al.}(1994)\citenamefont{Cichocki,
  Felderhof, Hinsen, Wajnryb, and Blawzdziewicz}}]{cichocki_friction_1994}
\bibinfo{author}{\bibfnamefont{B.}~\bibnamefont{Cichocki}},
  \bibinfo{author}{\bibfnamefont{B.~U.} \bibnamefont{Felderhof}},
  \bibinfo{author}{\bibfnamefont{K.}~\bibnamefont{Hinsen}},
  \bibinfo{author}{\bibfnamefont{E.}~\bibnamefont{Wajnryb}}, \bibnamefont{and}
  \bibinfo{author}{\bibfnamefont{J.}~\bibnamefont{Blawzdziewicz}},
  \bibinfo{journal}{The Journal of Chemical Physics}
  \textbf{\bibinfo{volume}{100}}, \bibinfo{pages}{3780} (\bibinfo{year}{1994}).

\bibitem[{\citenamefont{Leonardo et~al.}(2007)\citenamefont{Leonardo, Keen,
  Leach, Saunter, Love, Ruocco, and Padgett}}]{di_leonardo_eigenmodes_2007}
\bibinfo{author}{\bibfnamefont{R.~D.} \bibnamefont{Leonardo}},
  \bibinfo{author}{\bibfnamefont{S.}~\bibnamefont{Keen}},
  \bibinfo{author}{\bibfnamefont{J.}~\bibnamefont{Leach}},
  \bibinfo{author}{\bibfnamefont{C.~D.} \bibnamefont{Saunter}},
  \bibinfo{author}{\bibfnamefont{G.~D.} \bibnamefont{Love}},
  \bibinfo{author}{\bibfnamefont{G.}~\bibnamefont{Ruocco}}, \bibnamefont{and}
  \bibinfo{author}{\bibfnamefont{M.~J.} \bibnamefont{Padgett}},
  \bibinfo{journal}{Physical Review E.} \textbf{\bibinfo{volume}{76}}, \bibinfo{pages}{061402}
  (\bibinfo{year}{2007}).

\end{thebibliography}

\end{document}